\documentclass{article}
\usepackage{spconf,amsmath,graphicx}
\usepackage{epsfig}
\usepackage{xspace}
\usepackage{srcltx}
\usepackage{caption}
\usepackage{subcaption}
\usepackage[export]{adjustbox}
\usepackage{multirow}
\usepackage{booktabs}
\usepackage{textcomp}

\usepackage{color}
\usepackage[dvipsnames]{xcolor}

\newcommand{\CMT}[1]{{}}

\newcommand{\tbh}[1]{\textbf{#1}}

\title{Multi-Dialect Speech Recognition With a Single \\ Sequence-to-Sequence Model}
\name{Bo Li, Tara N. Sainath, Khe Chai Sim, Michiel Bacchiani,}
\secondlinename{Eugene Weinstein, Patrick Nguyen, Zhifeng Chen, Yonghui Wu, Kanishka Rao}
\address{Google Inc., USA \\ 
\fontsize{9}{9}\selectfont\ttfamily\upshape
\{boboli,tsainath,khechai,michiel,weinstein,drpng,zhifengc,yonghui,kanishkarao\}@google.com}

\begin{document}
\ninept
\maketitle
\begin{abstract}
  Sequence-to-sequence models provide a simple and elegant solution for building speech recognition systems by folding separate components of a typical system, namely acoustic (AM), pronunciation (PM) and language (LM) models into a single neural network. In this work, we look at one such sequence-to-sequence model, namely listen, attend and spell (LAS) \cite{ChanJaitlyLeEtAl15}, and explore the possibility of training a single model to serve different English dialects, which simplifies the process of training multi-dialect systems without the need for separate AM, PM and LMs for each dialect. We show that simply pooling the data from all dialects into one LAS model falls behind the performance of a model fine-tuned on each dialect. We then look at incorporating dialect-specific information into the model, both by modifying the training targets by inserting the dialect symbol at the end of the original grapheme sequence and also feeding a 1-hot representation of the dialect information into all layers of the model. Experimental results on seven English dialects show that our proposed system is effective in modeling dialect variations within a single LAS model, outperforming a LAS model trained individually on each of the seven dialects by 3.1\texttildelow 16.5\% relative.
\end{abstract}


\section{Introduction \label{sec:introduction}}

Dialects are defined as variations of the same language, specific to geographical regions or social groups. Although they share many similarities, there are usually large differences at several linguistic levels; amongst others: phonological, grammatical, orthographic (e.g., ``color'' {\it vs.} ``colour'') and very often different vocabularies. As a result, automatic speech recognition (ASR) systems trained or tuned for one specific dialect perform poorly when tested on another dialect of the same language \cite{biadsy2012google}. In addition, systems simultaneously trained on many dialects fail to generalize well for each individual dialect \cite{elfeky2016towards}. Inevitably, multi-dialect languages pose a challenge to ASR systems. If enough data exists for each dialect, a common practice is to treat each dialect independently \cite{biadsy2012google, van2001recognizing, elfeky2015multi}. Alternatively, in cases where dialects are resource-scarce, these models are boosted with data from other dialects~\cite{elfeky2016towards, lin2009study}. In the past, there have been many attempts to build multi-dialect/language systems. The usual approach has been to define a common set of universal phone models \cite{wang2002towards, lin2008learning, vu2014multilingual} with appropriate parameter sharing \cite{lin2009study} and train it on data from many languages with eventual adaptation on the data from the language of interest \cite{thomas2010cross,heigold2013multilingual,rao2017multi, byrne2000towards}. \cite{heigold2013multilingual,ghoshal2013multilingual} developed similar neural network models with language independent feature extraction and language dependent phonetic classifiers. \cite{rao2017multi} further investigated a grapheme-based multi-dialect model with dialect-dependent phoneme recognition as secondary tasks. Adaptation techniques such as MLLR and MAP are commonly used for Gaussian mixture model based systems \cite{byrne2000towards}; but for neural network based models, adaptation by continued training on dialect-specific data works well \cite{rao2017multi}.

We believe it has been challenging to build a universal multi-dialect model for conventional ASR systems because these models still require a separate pronunciation and language model (PM and LM) per dialect, which are trained independently from the multi-dialect acoustic model (AM). Therefore, if the AM predicts an incorrect set of sub-word units from the wrong dialect, errors get propagated to the PM and LM.


Sequence-to-sequence models provide a simple and elegant solution to the ASR problem by learning and optimizing a single neural network for the AM, PM and LM \cite{graves2012sequence, graves2014towards, chorowski2015attention, zhang2017very, lu2016training, prabhavalkar2017comparison}. We believe such a model is very attractive to look at for building a single multi-dialect system specifically for this reason. Training a multi-dialect sequence-to-sequence model is simple, as it requires simply pooling all the grapheme symbols together. In addition, the AM, PM and LM variations are jointly modeled across dialects \cite{kanthak2003multilingual}. The simplicity and joint optimization make it a good candidate for training multi-dialect systems.

In this work, we adopt the attention-based sequence-to-sequence model, namely listen, attend and spell (LAS) \cite{ChanJaitlyLeEtAl15}, for multi-dialect modeling. It has shown good performance compared to other sequence-to-sequence models for single dialect tasks \cite{prabhavalkar2017comparison}. Our goal here is to build a single LAS model for seven English dialects. We start by simply pooling all the data together. For English the grapheme set is shared across dialects, nothing needs to be modified for the output. Although this model gives acceptable performance on each dialect, it falls behind the ones independently fine-tuned on each dialect.

In the literature, adaptation methods \cite{diakoloukas1997development, huang2014multi} are typically used to integrate dialect-specific information into the system.
We hypothesize that by explicitly providing dialect information to the LAS model we should be able to bridge the gap between the dialect-independent and dialect-dependent models.
Firstly, we use dialect information in the output by introducing an artificial token into the grapheme sequence \cite{johnson2016google}.
The LAS model needs to learn both grapheme prediction and dialect classification. Secondly, we feed the dialect information as input vectors to the system. It can be either used as an extra information vector appended to the inputs of each layer or as weight coefficients for cluster adaptive training (CAT) \cite{tan2016cluster}. Our experimental results show that using dialect information can elevate the performance of multi-dialect LAS system to outperform dialect-dependent ones. The proposed system has several attractive benefits: 1) simplicity: no changes are required for the model and scaling to more dialects is trivial - by simply adding more data; 2) improvement for low-resource dialects: in the multi-dialect system, the majority of the parameters are implicitly shared by all the dialects, which forces the model to generalize across dialects during training. It is observed that the recognition quality on the low resource dialect is significantly improved.


\section{Multi-Dialect LAS Model}
\label{sec:arch}


The LAS \cite{ChanJaitlyLeEtAl15} model consists of an \emph{encoder} (akin to an acoustic model), a \emph{decoder} (like a language model) and an \emph{attention} model which learns an alignment between the encoder and decoder outputs. The encoder is normally a stack of recurrent layers; in this study we use 5 layers of unidirectional long short term memory (LSTM) \cite{hochreiter1997long}. The decoder is a neural language model, consisting of 2 LSTM layers. The attention module takes in the decoder's lowest layer's state vector from the previous time step and estimates attention weights for each frame in the encoder output in order to compute a single context vector. The context vector is then input into the decoder network, along with the previously predicted label from the decoder to generate logits from the final layer in the decoder. Finally, these logits are input into a softmax layer, which outputs a probability distribution over the label inventory (i.e., graphemes), conditioned on all previous predictions. In conventional LAS models, the label inventory is augmented with two special symbols, {\tt <sos>}, which is input to the decoder at the first time-step, and {\tt <eos>}, which indicates the end of a sentence. During inference, the label prediction process terminates when the {\tt <eos>} label is generated.

The baseline multi-dialect LAS system is built by simply pooling all the data together. The output targets consist of 75 graphemes for English, which are shared across dialects. In the following section, we describe how we can improve the baseline multi-dialect LAS model by providing dialect information. We assume this information is known in advance or can be easily obtained \cite{elfeky2015multi}. We believe explicitly providing such dialect information would be helpful to improve the performance of the multi-dialect LAS model. We discuss three ways of passing the dialect information into the LAS model, namely feeding it as output targets, as input vectors, or directly factoring the encoder layers based on the dialect.

\subsection{Dialect Information as Output Targets}

A common way to make the LAS model aware of the dialect is through multi-task learning \cite{seltzer2013multi}. We can add an extra dialect classification cost to the training and regularize the model to learn hidden representations that are effective for both dialect classification and grapheme prediction. However, this requires having two separate objective functions that are weighted, and deciding the optimal weight for each task is a parameter that needs to be swept \cite{seltzer2013multi}.

A simpler approach, similar to \cite{johnson2016google}, is to expand the label inventory of our LAS model to include a list of special symbols, each corresponding to a dialect. For example, when including the British English, we add the symbol {\tt <en-gb>} into the label inventory. In \cite{johnson2016google}, the special symbol is added to the beginning of the target label sequence. For example, for a British accented speech utterance of ``hello world'', the conventional LAS model uses ``{\tt <sos> h e l l o~\textvisiblespace~w o r l d <eos>}'' as the output targets; in the new setup the output target is ``{\tt <sos> <en-gb> h e l l o~\textvisiblespace~w o r l d <eos>}''. The model needs to figure out which dialect the input speech is first before making any grapheme prediction.

In LAS, each label prediction is dependent on the history. Adding the dialect symbol at the beginning, we implicitly incur the dependency of the grapheme prediction on the dialect classification. When the model makes errors in dialect classification, it may hurt the grapheme recognition performance. In this study, we assume the correct dialect information is always available. Hence, we explore inserting the dialect symbol at the end of the label sequence as well. For the example utterance, the target sequence now become ``{\tt <sos> h e l l o~\textvisiblespace~w o r l d <en-gb> <eos>}''. By inserting the dialect symbol at the end, the model still needs to learn a shared representation but avoids the unnecessary dependency and is less sensitive to dialect classification errors.



\subsection{Dialect Information as Input Vectors}

Another way of providing dialect information is to pass this information as an additional feature \cite{abdel2013fast}. To convert the categorical dialect information into a real-valued feature vector, we investigate the use of 1-hot vectors, whose values are all `0' except for one `1' at the index corresponding to the given dialect, and data-driven embedding vectors whose values are learned during training.
These dialect vectors can be appended to different layers in the LAS model. At each layer the dialect vectors are linearly transformed by the weight matrices and added to the original hidden activations before the nonlinearity. This effectively enables the model to learn dialect-dependent biases. We are mainly interested in two configurations: 1) adding it to the encoder layers, which effectively provides dialect information to help model the acoustic variations across dialects; and 2) appending it to the decoder layers, which models dialect-specific language model variations. Ultimately, we can also combine the two by feeding dialect vectors into both the encoder and the decoder.

\subsection{Dialect Information as Cluster Coefficients}

Another approach to modeling variations in the speech signal (for example dialects) which has been explored for conventional models is cluster adaptive training (CAT) \cite{tan2016cluster}. We can treat each dialect as a separate cluster and use 1-hot dialect vectors to switch clusters; alternatively, we can use data-driven dialect embedding vectors as weights to combine clusters. A drawback of the CAT approach is that it adds extra network layers, which typically adds more parameters to the LAS model.
In this study, our goal is to maintain simplicity of the LAS model and limit the increase in model parameters. Thus, we only tested a simple CAT setup for the encoder of the LAS model to compare with the input vector approaches discussed in the previous sections. 
Specifically, we use a few clusters to compensate activation offsets of the 4th LSTM layer based on the shared representation learned by the 1st LSTM layer, to account for the dialect differences. For each cluster, a single layer 128D LSTM is used with output projection to match the dimension of the 4th LSTM layer. The weighted sum of all the CAT bases using dialect vectors as interpolation weights is added back to the 4th LSTM layer's outputs, which are then fed to the last encoder layer.


\section{Experimental Details}
\label{sec:exps}

Our experiments are conducted on about 40K hours of noisy training data consisting of 35M English utterances. The training utterances are anonymized and hand-transcribed, and are representative of Google's voice search traffic. It includes speech from 7 different dialects, namely America (US), India (IN), Britain (GB), South Africa (ZA), Australia (AU), Nigeria \& Ghana (NG) and Kenya (KE). The amount of dialect-specific data can be found in Table \ref{table:data_train}. The training data is created by artificially corrupting clean utterances using a room simulator, adding varying degrees of noise and reverberation such that the overall SNR is between 0 and 20dB \cite{kim2017generation}. The noise sources are from YouTube and daily life noisy environmental recordings. We report results on dialect-specific test sets, each contains roughly 10K anonymized, hand-transcribed utterances from Google's voice search traffic without overlapping with the training data. This amounts to roughly 11 hours of test data per dialect. All experiments use 80-dimensional log-mel features, computed with a 25ms window and shifted every 10ms. Similar to \cite{sak2015fast, pundak2016lower}, at the current frame, $t$, these features are stacked with 3 frames to the left and downsampled to a 30ms frame rate. In the baseline LAS model, the encoder network architecture consists of 5 unidirectional 1024D LSTM \cite{hochreiter1997long} layers. Additive attention \cite{ChanJaitlyLeEtAl15} is used for all experiments. The decoder network is a 2-layer 1024D unidirectional LSTM.  All networks are trained to predict graphemes, which have 75 symbols in total. The model has a total number of 60.6M parameters.
All networks are trained with the cross-entropy criterion, using asynchronous stochastic gradient descent (ASGD) optimization \cite{dean2012large}, in TensorFlow \cite{abadi2016tensorflow}.
The training terminates when the change of WERs on a dev set is less than a given threshold for certain number of steps.


\begin{table}[ht]
\caption{Number of utterances per dialect for training (M for million) and testing (K for thousand).}
\vspace{-0.1in}
\centering
\begin{tabular}{l||r|r|r|r|r|r|r}
\toprule
\tbh{Dialect}  & \tbh{US} & \tbh{IN} & \tbh{GB} & \tbh{ZA} & \tbh{AU} & \tbh{NG} & \tbh{KE} \\
\midrule
\tbh{Train(M)} & 13.7     & 8.6      & 4.8      & 2.4      & 2.4      & 2.1      & 1.4 \\
\midrule
\tbh{Test(K)}  & 12.9     & 14.5     & 11.1     & 11.7     & 11.7     & 9.8      & 9.2   \\
\bottomrule
\end{tabular}
\label{table:data_train}
\vspace{-0.1in}
\end{table}

\section{Results}
\label{sec:results}


\subsection{Pooling All Data}
Firstly, we build a single grapheme LAS model on all the data together ({\tt S1} in Table \ref{tbl:baseline}). For comparison, we also build a set of dialect-dependent models. Due to the large variations in the amount of data we have for each dialect, a lot of tuning is required to find the best model setup from scratch for each dialect. For the sake of simplicity, we take the joint model as the starting point and retraining the same architecture for each dialect independently ({\tt S2} in Table \ref{tbl:baseline}). Instead of updating only the output layers \cite{heigold2013multilingual,ghoshal2013multilingual}, we find reestimating all the parameters work better. To compensate for the extra training time the fine-tuning brings in, we also keep the baseline model training for similar extra number of steps; we do not find to improve the WER. Comparing the dialect-independent model ({\tt S1}) with the dialect-dependent ones ({\tt S2}), simply pooling the data together gives acceptable recognition performance, but having a language-specific model by fine-tuning still achieves better performance.

\begin{table}[ht]
\caption{WER (\%) of dialect-independent ({\tt S1}) and dialect-dependent ({\tt S2}) LAS models.}
\vspace{-0.1in}
\centering
\begin{tabular}{l||r|r|r|r|r|r|r}
\toprule
\tbh{Dialect}  & \tbh{US} & \tbh{IN} & \tbh{GB} & \tbh{ZA} & \tbh{AU} & \tbh{NG} & \tbh{KE} \\
\midrule
\tbh{\tt S1}       & 10.6     & 18.3      & 12.9     & 12.7     & 12.8     & 33.4     & 19.2 \\
\midrule
\tbh{\tt S2}       &{\bf 9.7} &{\bf 16.2} &{\bf 12.7}&{\bf 11.0}&{\bf 12.1}& 33.4     &{\bf 19.0} \\
\bottomrule
\end{tabular}
\vspace{-0.1in}
\label{tbl:baseline}
\end{table}

\subsection{Using Dialect-Specific Information}
Our next set of experiments look at using dialect information to see if we can have a joint multi-dialect model improve performance over the dialect-specific models ({\tt S2}) in Table \ref{tbl:baseline}.


\subsubsection{Results using Dialect Information as Output Targets}

We first add the dialect information into the target sequence.
Two setups are explored, namely adding at the beginning ({\tt S3}) and adding at the end ({\tt S4}). The results are presented in Table \ref{tbl:did}. Inserting the dialect symbol at the end of the label sequence is much better than at the beginning, which eliminates the dependency of grapheme prediction on the erroneous dialect classification. {\tt S4} is more preferable and outperforms the dialect-dependent model ({\tt S2}) on all the dialects except for IN and ZA.


\begin{table}[ht]
\caption{WER (\%) of inserting dialect information at the beginning ({\tt S3}) or at the end ({\tt S4}) of the grapheme sequence.}
\vspace{-0.1in}
\centering
\begin{tabular}{l||r|r|r|r|r|r|r}
\toprule
\tbh{Dialect}  & \tbh{US} & \tbh{IN} & \tbh{GB} & \tbh{ZA} & \tbh{AU} & \tbh{NG} & \tbh{KE} \\
\midrule
\tbh{\tt S2}       & 9.7      &{\bf 16.2}& 12.7     &{\bf 11.0}& 12.1     & 33.4     & 19.0 \\
\midrule
\tbh{\tt S3}       & 9.9      & 16.6     & 12.3     & 11.6     & 12.2     & 33.6     & 18.7 \\
\tbh{\tt S4}       &{\bf 9.4} & 16.5     &{\bf 11.6}&{\bf 11.0}&{\bf 11.9}&{\bf 32.0}&{\bf 17.9}\\
\bottomrule
\end{tabular}
\vspace{-0.1in}
\label{tbl:did}
\end{table}

\subsubsection{Results using Dialect Information as Input Vectors}

\begin{table}[ht]
\caption{WER (\%) of feeding the dialect information into the LAS model's encoder ({\tt S5}), decoder ({\tt S6}) and both ({\tt S7}). The dialect information is converted into an 8D vector using either 1-hot representation ({\tt 1hot}) or learned embedding ({\tt emb}).}
\vspace{-0.1in}
\centering
\begin{tabular}{l||r|r|r|r|r|r|r}
\toprule
\tbh{Dialect}  & \tbh{US} & \tbh{IN} & \tbh{GB} & \tbh{ZA} & \tbh{AU} & \tbh{NG} & \tbh{KE} \\
\midrule
\tbh{\tt S2}       & 9.7      & 16.2     & 12.7     & 11.0     & 12.1     & 33.4     & 19.0 \\
\midrule
\tbh{\tt S5(1hot)}& 9.6      & 16.4     & 11.8     & 10.6     & 10.7     & 31.6     & 18.1 \\
\tbh{\tt S5(emb)} & 9.6      & 16.7     & 12.0     & 10.6     & 10.8     & 32.5     & 18.5 \\
\midrule
\tbh{\tt S6(1hot)}& 9.4      & 16.2     & 11.3     & 10.8     & 10.9     & 32.8     & 18.0 \\
\tbh{\tt S6(emb)} & 9.4      & 16.2     &{\bf 11.2}& 10.6     & 11.1     & 32.9     & 18.0 \\
\midrule
\tbh{\tt S7(1hot)}&{\bf 9.1} &{\bf 15.7}& 11.5     &{\bf 10.0}&{\bf 10.1}&{\bf 31.3}&{\bf 17.4}\\
\bottomrule
\end{tabular}
\vspace{-0.1in}
\label{tbl:vector}
\end{table}


Next we experiment with directly feeding the dialect information into different layers of the LAS model. The dialect information is converted into an 8D vector using either 1-hot representation or an embedding vector learned during training. This vector is then appended to both the inputs and hidden activations. We compare the usefulness of this dialect vector to the LAS encoder and decoder. From Table \ref{tbl:vector}, feeding it to encoder ({\tt S5}) gives gains on dialects with less data (namely GB, ZA, AU, NG and KE) and has comparable performance on US, but is still a bit worse on IN compared to the fine-tuned dialect-dependent models ({\tt S2}). Similarly, we pass the dialect vector (using both 1-hot and learned embedding) to the decoder of LAS ({\tt S6}). Table \ref{tbl:vector} shows that in this way the single multi-dialect LAS model outperforms the individually fine-tuned dialect-dependent models on all dialects except for IN, for which it obtains the same performance.

Comparing the use of 1-hot representation and learned embedding, we do not observe big differences for both the encoder and decoder. It is most likely the small dimensionality of the vectors used ({\it i.e.} 8D) that is insufficient to suggest any preference between the 1-hot representation and the learned embedding.
In future, when scaling up to more dialects/languages, using embedding vectors instead of 1-hot to represent a larger set of dialects/languages could be a more economical way.

\begin{figure}[t!]
\vspace{-0.1in}
\centering
  \begin{subfigure}[b]{0.23\textwidth}
        \includegraphics[width=\textwidth]{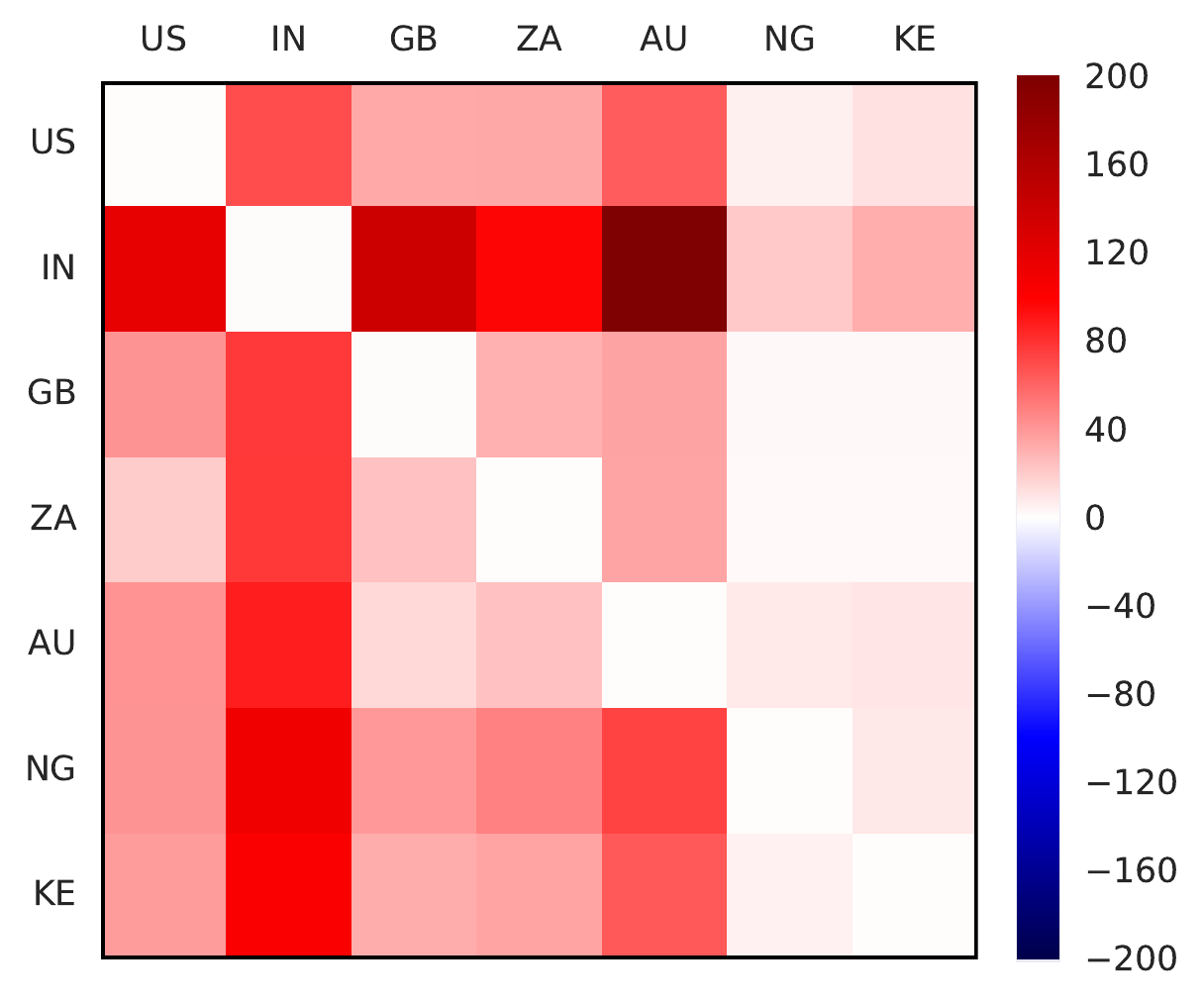}
        \caption{Encoder ({\tt S5(1hot)})}
        \label{fig:analysis_enc}
    \end{subfigure}
    ~ 
    \begin{subfigure}[b]{0.23\textwidth}
        \includegraphics[width=\textwidth]{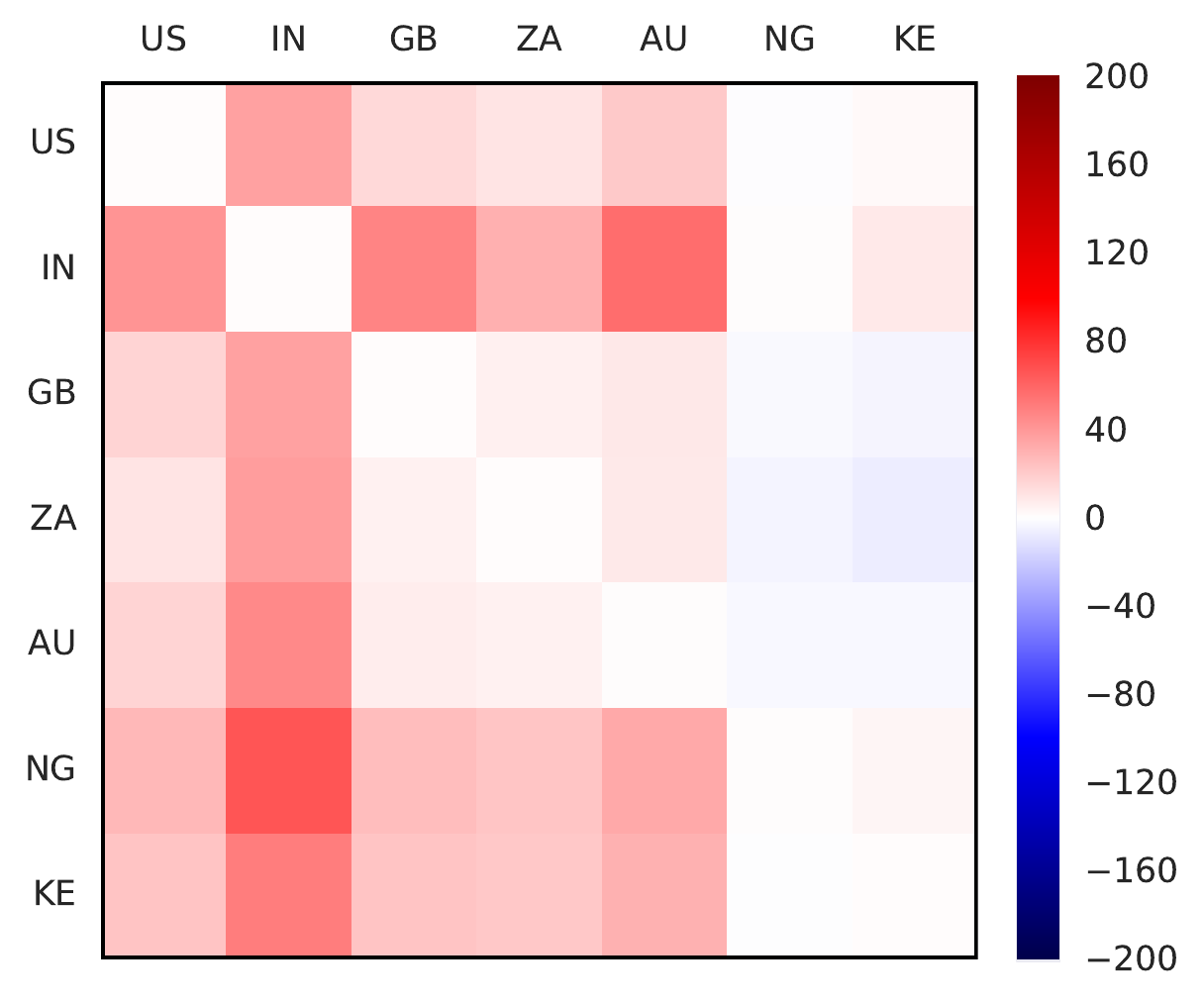}
        \caption{Decoder ({\tt S6(1hot)})}
        \label{fig:analysis_dec}
    \end{subfigure}
 \vspace{-0.1in}
  \caption{Relative WER changes when feeding in wrong dialect vectors (rows) to the encoder or decoder for each test set (columns). The red and blue colors indicate the relative increase and decrease of WERs respectively and the white color means no change in WERs.}
  \label{fig:analysis}
  \vspace{-0.2in}
\end{figure}

Feeding dialect vectors into different layers effectively enables the model to explicitly learn dialect-dependent biases. For the encoder, these biases would help capture dialect-specific acoustic variations; while in the decoder, they can potentially address the language model variations. Experimental results suggest that these simple biases indeed help the multi-dialect LAS model. To understand their effects, we test system {\tt S5(1hot)} and {\tt S6(1hot)} with mismatched dialect vector on each test set. The relative WER changes are depicted in Figure \ref{fig:analysis}. Each row represents the dialect vector fed into the model and each column corresponds to a dialect-specific test set. The white diagonal blocks are the ``correct'' setups, where we feed in the correct dialect vector on each test set. The red and blue colors represent the relative increase and decrease of WERs respectively. The darker the color is, the larger the change is. Comparing the effect on encoder and decoder, wrong dialect vectors degrades more on encoders, suggesting more acoustic variations across dialects than language model differences. Across different dialects, IN seems to have the most distinguishable characteristics. NG and KE, the two smallest dialects in this study, benefit more from the sharing of parameters as the performance varies little with different dialect vectors. This suggests the proposed model is capable of handling the unbalanced dialect data properly, learning strong dialect-dependent biases when there's enough data and sticking to the shared model otherwise. Another interesting observation is that, for these two dialects, feeding dialect vectors from ZA is slightly better than using their own. This suggests in future pooling similar dialects with less data may give better performance.

One evidence that the model successfully learns dialect-specific lexicons is ``color'' in US {\it vs.} ``colour'' in GB. On the GB test set, the system without any explicit dialect information ({\tt S1}) and the one feeding it only to encoder layers ({\tt S5}) generate recognition hypotheses with both ``color'' and ``colour'' although ``color'' appears much less frequently. However, for the model {\tt S6}, where the dialect information is directly fed into decoder layers, only ``colour'' appears; moreover, if we feed in the dialect vector for US to {\tt S6} on the GB test set, the model successfully switches all the ``colour'' predictions to ``color''. Similar observations are found for ``labor'' {\it vs.} ``labour'', ``center'' {\it vs.} ``centre'' etc.


Next, we feed the 1-hot dialect vector into all the layers of the LAS model ({\tt S7}). Experimental results (Table \ref{tbl:vector}) show that this system outperforms the dialect-dependent models on all the test sets, with the largest gains on AU (16.5\% relative WER reduction).

\subsubsection{Results using Information as Cluster Coefficients}
In the literature, instead of directly feeding the dialect vector as inputs to learn a simple bias, it can also be used as a cluster coefficient vector to combine multiple clusters and learn more complex mapping functions. For comparisons, we implement a simple CAT system ({\tt S8}) only for the encoder.
Experimental results in Table \ref{tbl:cat} show that unlike directly feeding dialect vectors as inputs, CAT favors more learned embeddings ({\tt S8(emb)}), which encourages more parameter sharing across dialects.
In addition, comparing this to directly using dialect vectors ({\tt S5(1hot)}) for the encoder, CAT ({\tt S8(emb)}) is more effective on US and IN and similar on other dialects. However, in terms of model size, comparing to the baseline model ({\tt S1}), {\tt S5(1hot)} only increases by 160K parameters, while {\tt S8(emb)} adds around 3M extra. We will leave a more thorough study of CAT for future work.

\begin{table}[ht]
\caption{WER (\%) of a CAT encoder LAS system ({\tt S8}) with 1-hot ({\tt 1hot}) and learned embedding ({\tt emb}) dialect vector.}
\vspace{-0.1in}
\centering
\begin{tabular}{l||r|r|r|r|r|r|r}
\toprule
\tbh{Dialect}  & \tbh{US} & \tbh{IN} & \tbh{GB} & \tbh{ZA} & \tbh{AU} & \tbh{NG} & \tbh{KE} \\
\midrule
\tbh{\tt S2}       & 9.7      & 16.2     & 12.7     & 11.0     & 12.1     & 33.4     & 19.0 \\
\midrule
\tbh{\tt S5(1hot)}& 9.6      & 16.4     & 11.8     & 10.6     & 10.7     &{\bf 31.6}&{\bf 18.1}\\
\midrule
\tbh{\tt S8(1hot)}& 9.9      & 17.0     & 12.1     & 11.0     & 11.6     & 32.5     & 18.3 \\
\tbh{\tt S8(emb)} &{\bf 9.4} &{\bf 16.1}&{\bf 11.7}&{\bf 10.6}&{\bf 10.6}& 32.9     &{\bf 18.1}\\
\bottomrule
\end{tabular}
\vspace{-0.1in}
\label{tbl:cat}
\end{table}

\subsection{Combining Adaptation Strategies}
Lastly, we integrate the joint dialect identification ({\tt S4}) and the use of dialect vectors ({\tt S7(1hot)}) into a single system ({\tt S9}). The performance of this combined multi-dialect LAS system is presented in Table \ref{tbl:final}. It works much better than doing joint dialect identification ({\tt S4}) alone, but has similar performance to the one uses dialect vectors ({\tt S7(1hot)}). This is because when feeding in dialect vectors into the LAS model, especially in the decoder layers, the model is already doing a very good job in predicting the dialect. Specifically, the dialect prediction error for {\tt S9} on the dev set during training is less than 0.001\% compared to {\tt S4}'s 5\%. Overall, our best multi-dialect system ({\tt S7(1hot)}) outperforms dialect-specific models and achieves 3.1\texttildelow 16.5\% WER reductions across dialects.

\begin{table}[ht]
\caption{WER (\%) of the combined multi-dialect LAS system ({\tt S9}).}
\vspace{-0.1in}
\centering
\begin{tabular}{l||r|r|r|r|r|r|r}
\toprule
\tbh{Dialect}  & \tbh{US} & \tbh{IN} & \tbh{GB} & \tbh{ZA} & \tbh{AU} & \tbh{NG} & \tbh{KE} \\
\midrule
\tbh{\tt S2}       & 9.7      & 16.2     & 12.7     & 11.0     & 12.1     & 33.4     & 19.0 \\
\midrule
\tbh{\tt S4}       & 9.4      & 16.5     & 11.6     & 11.0     & 11.9     & 32.0     & 17.9 \\
\tbh{\tt S7(1hot)} &{\bf 9.1} &{\bf 15.7}& 11.5     & 10.0     &{\bf 10.1}&{\bf 31.3}&{\bf 17.4}\\
\midrule
\tbh{\tt S9}       &{\bf 9.1} & 16.0     &{\bf 11.4}&{\bf 9.9} & 10.3     & 31.4     & 17.5 \\
\bottomrule
\end{tabular}
\vspace{-0.1in}
\label{tbl:final}
\end{table}

\section{Conclusions}
\label{sec:concl}

In this study, we explored a multi-dialect end-to-end LAS system trained on 7 English dialects. The model utilizes a 1-hot dialect vector at each layer of the LAS encoder and decoder to learn dialect-specific biases. It is optimized to predict the grapheme sequence appended with the dialect name as the last symbol, which effectively forces the model to learn shared hidden representations that are suitable for both grapheme prediction and dialect classification. Experimental results show that feeding a 1-hot dialect vector is very effective in boosting the performance of a multi-dialect LAS system, and allows it to outperform a LAS model trained on each individual language. Furthermore, we also find that using CAT could potentially be more powerful in modeling dialect variations though at a cost of increased parameters, which will be addressed in future work.


\vfill\pagebreak

\bibliographystyle{IEEEbib}
\bibliography{refs}

\end{document}